\documentclass{iopart}
\usepackage{graphicx} 
\usepackage{braket}
\usepackage{amssymb}
\usepackage{pstricks}
\usepackage{bbm}

\newcommand{\be}{\begin{equation}}
\newcommand{\ee}{\end{equation}}
\newcommand{\ben}{\begin{eqnarray}}
\newcommand{\een}{\end{eqnarray}}
\newcommand{\bes}{\begin{subequations}}
\newcommand{\ees}{\end{subequations}}
\newcommand{\bF}{\begin{figure}}
\newcommand{\eF}{\end{figure}}

\def\ket#1{ | #1 \rangle}

\newcommand{\proj}[1]{\mbox{$|#1\rangle \!\langle #1 |$}}

\newcommand{\adagg}{\hat{a}^{\dagger}}
\newcommand{\bdagg}{\hat{b}^{\dagger}}

\begin{document}

\title{Integrated photonic sensing}

\author{N~Thomas-Peter, N~K~Langford, A~Datta, L~Zhang, B~J~Smith, J~B~Spring, B~J~Metcalf, H~B~Coldenstrodt-Ronge, M~Hu, J~Nunn and I~A~Walmsley} 
\address{Clarendon Laboratory, University of Oxford, Parks Road, Oxford, UK, OX1 3PU}
\ead{Email: n.thomas-peter1@physics.ox.ac.uk}

\begin{abstract}
Loss is a critical roadblock to achieving photonic quantum-enhanced technologies.  We explore a modular platform for implementing integrated photonics experiments and consider the effects of loss at different stages of these experiments, including state preparation, manipulation and measurement.  We frame our discussion mainly in the context of quantum sensing and focus particularly on the use of loss-tolerant Holland-Burnett states for optical phase estimation.  In particular, we discuss spontaneous four-wave mixing in standard birefringent fibre as a source of pure, heralded single photons and present methods of optimising such sources.  We also outline a route to programmable circuits which allow the control of photonic interactions even in the presence of fabrication imperfections and describe a ratiometric characterisation method for beam splitters which allows the characterisation of complex circuits without the need for full process tomography.  Finally, we present a framework for performing state tomography on heralded states using lossy measurement devices.  This is motivated by a calculation of the effects of fabrication imperfections on precision measurement using Holland-Burnett states. 
\end{abstract}
\submitto{New Journal of Physics}

\maketitle

\section{Introduction}
The field of integrated photonics is a promising area for the development of quantum-enhanced technologies with applications being pursued in communications, metrology, simulation and information processing~\cite{SmithBJ2009pip, MatthewsJCF2009mme, PeruzzoA2010qwc, SchreiberA2010pwl, PolitiA2008swq, PolitiA2009sqf}.  These quantum experiments can be broadly broken up into three stages:  state preparation, manipulation, and measurement.  Over the last decade or so, integrated approaches have made substantial advances in each of these three areas, particularly with regard to fibre- and waveguide-based nonclassical light sources~\cite{AndersonME1997psg, BanaszekK2001gcp, URenAB2004ecp, ChenJ2005tgf, XiaoyingL2005osp, SharpingJE2006gcp, Garay-PalmettK07ppp, AvenhausM2009fss, ChenJ2009vws, ClemmenS2009cwp, MosleyPJ2009dms, SmithBJ2009ppg, LevineZH2010hps, BranczykAM2010ogh, EcksteinA2011hes}, on-chip quantum circuits~\cite{HonjoT2004dqk, TakesueH2005g1b, PolitiA2008swq, SmithBJ2009pip, MarshallGD2009lww, MatthewsJCF2009mme, SansoniL2010pes, PeruzzoA2010mqi, WuB2010slc}, and integrated detection~\cite{BanaszekK2003pcl, AchillesD2006dlc, NatarajanCM2010oqw}.

Perhaps the greatest challenge facing photonic quantum applications is the issue of photon loss, which is generally present in all three of these stages.  Loss is particularly critical in quantum information processing (QIP) and quantum-enhanced metrology, where strict thresholds exist for the efficiencies required for a device to out-perform its classical counterpart~\cite{VarnavaM2006lto, VarnavaM2008hgm, Thomas-PeterNL2010rqs, DattaA2010qmi}.  The most significant contributions to loss in photonic experiments generally arise at the interfaces between different stages of the experiment, e.g., between sources, which may be either spatially or spectrally multimode, and circuits, which often rely critically on low-distinguishability quantum interference.  In this example, bulk optics experiments typically require the use of strong, inherently lossy spatial and spectral filtering to reach the necessary operation fidelities.

The key advantage of integrated optics in overcoming these losses is that it provides the potential for complete spatial and spectral control of the underlying optical field modes.  The tight confinement and guided nature of these modes provides exquisite control of their spatial properties, both in terms of their transverse profiles and paths of propagation.  This enables strong interactions between highly indistinguishable optical modes.  It is therefore possible to fabricate devices with precise, stable, and potentially complex circuit configurations which, because they are small and monolithic, are inherently stable without the need for complex stabilisation techniques and provide high-visibility interference~\cite{PittmanTB2003ecl, SmithBJ2009pip, MarshallGD2009lww, MatthewsJCF2009mme, LaingA2010hoq}.  This holds great promise for many photonic QIP applications, which can require complex circuits containing many nested interferometers~\cite{LanyonBP2007edc, WaltherP2005eoq, ZhangQ2006eqt, PrevedelR2007hlo, GaoWB2010erc}, by alleviating the impractical space and stability requirements facing bulk-optics implementations.  At the same time, by carefully managing the dispersion properties of integrated devices, either by natural means (material selection) or engineering design (e.g., poling or mode confinement), it is also possible to exercise fine control over the spectral properties of the optical modes.  Such control can be used, for example, to design spectrally pure, heralded single-photon sources~\cite{URenAB2004ecp, RaymerMG2005psw, ChristA2009psp, BranczykAM2010ogh, EcksteinA2011hes}, a critical requirement for scaleable photonic QIP systems.

The holy grail of integrated photonics is a single device containing an entire photonic quantum experiment, state preparation, manipulation, and measurement.  Such an approach, however, faces an enormous challenge; a fully integrated platform requires the marriage of many fundamentally different, often-incompatible technologies within a single material substrate.  For example, thermo-optic phase control~\cite{SmithBJ2009pip, MatthewsJCF2009mme} would clearly be impossible on a device which is being cooled to cryogenic temperatures (mK) in order to operate a superconducting TES (transition-edge sensor), which currently provide the leading performance in terms of both quantum efficiency and photon-number resolution.

An alternative approach, achievable within the current state of the art, is to employ a more modular solution, where each stage of the experiment is performed using the technology platform that is most suited to the task.  The main task of integration is then simply to ensure that the interface between platforms does not introduce an unacceptable level of loss.  This is not an unreasonable target, however, since the unprecedented spatial-mode control available in integrated architectures already provides the means for optimising the coupling efficiencies between platforms.  Such an approach also presents a strong advantage in terms of flexibility.  This cluster of technologies would become a platform on which large scale complex experiments could be performed, enabling progress in both practical applications of quantum enhanced technologies and fundamental research, and could include components for active phase control~\cite{MatthewsJCF2009mme, SmithBJ2009pip}, Bragg reflectors for frequency filtering and building cavities~\cite{RaymerMG2005psw}, and on-chip micro-fluidic capillaries for interfacing photonic systems with fluids or fluid suspensions~\cite{SchmidtH2005hw2}, to name just a few examples.

In this paper we will focus on the specific example of sensing in a real-world integrated photonic device via (linear) quantum interferometry as it provides a clear example of the stringent requirements placed upon the different modules.  We will follow the chain of state generation, manipulation, and detection, discussing in turn the key requirements for and issues surrounding each.  We will also introduce a technique for characterising optical elements in an integrated photonic device, as well as a framework which allows tomographic reconstruction of a heralded state, including all photon number subspaces.  

\section{Quantum sensing---an example context}
\begin{figure}[h]
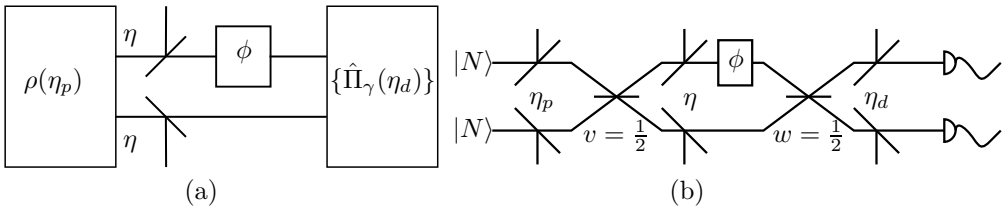

\begin{minipage}[b]{0.4\textwidth}
\centering
\def\svgwidth{1.1\textwidth}
\input{figure1a.tex}
(a)
\end{minipage}
\hspace{1cm}
\begin{minipage}[b]{0.4\textwidth}
\centering
\def\svgwidth{1.3\textwidth}
\input{figure1b.tex}
(b)
%\vspace{0.1cm}
\end{minipage}
\caption{Schematic of a typical metrology scenario. (a) General two mode interferometer with inefficient state preparation, two lossy modes, and inefficient detection.  (b) A detailed equivalent network involving a Mach-Zehnder Interferometer (MZI) with the losses specified by individual beam splitters of transmissivity $\eta_{p}$, $\eta$, and $\eta_{d}$.}
\label{fig:schemSetup}
\end{figure}

Quantum-enhanced metrology is concerned with a single task: preparing a quantum state which is sensitive to a parameter, $\phi$, and implementing a measurement on that state so that the uncertainty in the measurement of the parameter, $\Delta\phi$, is lower than the uncertainty that would be obtained using the same number of classical resources.  Typically, only the scaling behaviour of $\Delta\phi$ with the number of particles in the state, $N$, is considered since for large enough $N$ a class of states which provides only a constant factor improvement can always be beaten by one which has a more favourable scaling.  The tightest known limit on precision is given by the Heisenberg limit (HL), $\Delta\phi \geq \Delta\phi_{HL} = 1/\sqrt{\nu}N$, where $\nu$ is the number of experimental trials.

The situation we will consider in this paper is shown schematically in figure~\ref{fig:schemSetup}(a).  For any given state and measurement, the input state $\rho$ evolves to become $\rho(\phi)$ and a measurement is performed with outcomes represented by the operators $\{\hat{\Pi}_{\gamma}\}$.  Performing $\nu$ trials provides a data set and the precision with which $\phi$ can be estimated from this data set, $\Delta\phi$, is bounded by the Cram\'er-Rao bound (CRB), $\Delta\phi\geq \Delta\phi_{CRB}=1/\sqrt{\nu F(\phi)}$.  Here, $F(\phi)$ is the Fisher information (FI), a measure of the information which can be obtained about $\phi$ from the data set.  The CRB can be saturated for large $\nu$ by maximum likelihood estimation.  Interestingly, a measurement-independent version of the FI can be found by maximising over all physical measurements giving the quantum Fisher Information (QFI), $F_{Q}(\phi)$.  This, in turn leads to the lowest possible bound on the precision obtainable using a given state, the quantum Cram\'er-Rao bound (QCRB) $\Delta\phi\geq\Delta\phi_{QCRB}=1/\sqrt{\nu F_{Q}(\phi)}$~\cite{BraunsteinSL1994sdg}.

To judge whether or not a device could perform better than its classical counterpart, we must compare the QCRB of the prepared state to that of $N$ completely uncorrelated photons.  In the ideal situation of a perfectly transmissive device, this leads to the standard quantum limit $\Delta\phi \geq \Delta\phi_{SQL} = 1/\sqrt{\nu N}$.  In the presence of loss, however, this becomes the standard interferometric limit (SIL)~\cite{DornerU2009oqp}, $\Delta\phi \geq \Delta\phi_{SIL} = 1/\sqrt{\nu\eta\eta_{d}N}$ where $\eta$ and $\eta_{d}$ model transmission inside the interferometer, and detection efficiency respectively (see figure~\ref{fig:schemSetup}).  The SIL is unaffected by the parameter $\eta_{p}$ as a state which saturates the SIL, the phase-averaged coherent state, is unaffected by $\eta_{p}$ apart from a scaling in average photon number.  It is therefore considered that a phase-averaged coherent state with arbitrary average photon number can be prepared perfectly.  In general, however, all losses including $\eta_{p}$ play a crucial role in the QCRB for any given prepared state.  If $\Delta\phi_{QCRB} \leq \Delta\phi_{SIL}$ then the state which is prepared could perform better than its classical counterpart, but it will only do so if an appropriate measurement scheme is implemented.  If such a scheme is implemented, $\Delta\phi_{CRB} \leq \Delta\phi_{SIL}$ and the device truly does outperform its classical counterpart.

\section{State generation}
The most widely studied state for quantum-enhanced metrology is the N00N state~\cite{MitchellMW2004spm, EisenbergHS2005mpe, HigginsBL2007ehp, SmithBJ2009pip, MatthewsJCF2009mme, KimH2009tns, AfekI2010hsm}.  This path-entangled state is a superposition of $N$ (0) photons in one mode with 0 ($N$) photons in the other and, under perfect transmission, it saturates the HL.  It does, however, suffer from a major drawback in the lossy case.  When even a single photon is lost from the N00N state, the remaining state is entirely insensitive to the phase $\phi$.  Thus all photons must make it through the device and be detected in order for information to be gained about $\phi$,  an exponentially unlikely event for increasing $N$~\cite{Thomas-PeterNL2010rqs}.  Schemes for generating high-$N$ N00N states are also complex, often requiring many nonlinear elements~\cite{WaltherP2007hgm} or carefully aligned cavities~\cite{McCuskerKT2009eoq}.  Optimally loss-tolerant $N$-photon states have also been studied~\cite{DornerU2009oqp, KacprowiczM2010eqe}, but preparation strategies and detection schemes for these states have not yet been found and knowledge of the exact channel loss would also be required.

By contrast to N00N states and optimal states, Holland-Burnett (HB) or twin-Fock states~\cite{HollandMJ1993ido} provide a much simpler route to quantum-enhanced metrology whilst retaining near-optimal loss tolerance.  An HB($N$) state is prepared by interfering two $N$-photon Fock states at a 50:50 beam splitter (see figure~\ref{fig:schemSetup}b).  The resulting state retains phase sensitivity under loss both before and after this beam splitter and hence does not need perfect transmission to be useful.  As well as this, the QFI for the HB($N$) state is attained through the use of photon-number-resolving detectors, regardless of the loss parameters~\cite{DattaA2010qmi}.  The requisite Fock states can be prepared by heralding $N$ photons from two two-mode photon-number-correlated sources, for example based on spontaneous parametric downconversion (SPDC) or spontaneous four-wave mixing (SFWM).  Recent work~\cite{DattaA2010qmi} has put bounds on the $\eta_{p}$, $\eta$, and $\eta_{d}$ that are necessary to beat the classical limit, showing that for $\eta_{d}=0.6$ and $\eta=0.95$, a state preparation efficiency of $\eta_{p}\geq0.91$ is required.  Even for the best currently available detectors with $\eta_{d}=0.98$, the requirement on $\eta_{p}$ is relaxed only to $\eta_{p}\geq0.71$.  This puts extremely challenging bounds on the current state of the art for quantum sources.

The key ingredient for generating HB($N$) states is the creation of $N$-photon Fock states with sufficient purity to achieve the quality of quantum interference required for high-fidelity state production.  More generally, if these sources are to be useful in a scaleable way for integrated photonics quantum technologies, they will need to be either on-demand or heralded.  A heralded source can be  converted into an on-demand source using a controllable quantum memory with a high time-bandwidth product~\cite{NunnJ2007mbs, ReimKF2010tho, ReimKF2010sqm}.  The important characteristics of such sources are heralding or preparation efficiency, purity, and indistinguishability of the photons produced (in polarization, spatial and temporal degrees of freedom).  In addition, in order to avoid excess loss at the interface with the quantum circuit, the photon mode structure should be compatible with integrated devices, particularly in terms of spatial overlap, to ensure optimal performance.  If the photons produced by these sources do not adhere to strict requirements for these parameters, the performance of the quantum protocols rapidly decays~(e.g., \cite{RalphTC2002loc, BarbieriM2009pdo}).

One of the most common and most successful ways to produce heralded photonic quantum states makes use of nonlinear processes such as either spontaneous parametric down-conversion (SPDC) or spontaneous four-wave mixing (SFWM).  These are robust, room-temperature processes that can produce large numbers of photons for moderate pump powers.  Because of the strong photon-number correlations induced by the nonlinear interaction, they can be used to produce high-quality heralded Fock states.  Unfortunately, however, the generated state also contains terms with more than the desired number of photon pairs in a single pulse.  This ambiguity leads to dramatically reduced performance in many QIP applications if standard avalanche photodiodes (APDs) are used as heralding detectors, because they are unable to distinguish between different photon-number terms~\cite{BarbieriM2009pdo}.  In the ideal case, this problem can be largely mitigated by instead using photon-number-resolving heralding detectors~\cite{AchillesD2006dlc, AvenhausM2008pns}.  In the presence of loss, however, it becomes exacerbated and cannot simply be overcome by using a stronger pump to increase the count rate.  Loss acts to weaken the photon-number correlations in the generated state and therefore decrease the quality of interference exhibited by the heralded Fock states.  Furthermore, although increasing the pump does increase the signal strength, it can increase the strength of the noise-contributing higher-order terms by a larger proportion, because although they occur less frequently than the signal term, they are often more likely to be detected when they do occur.  Thus, quantum state production must be optimised to provide the highest possible heralding efficiency.

\begin{figure}[h!]
\begin{minipage}[b]{1\textwidth}
\begin{minipage}[b]{0.4\textwidth}
\centering
(a)
\includegraphics[width=\textwidth]{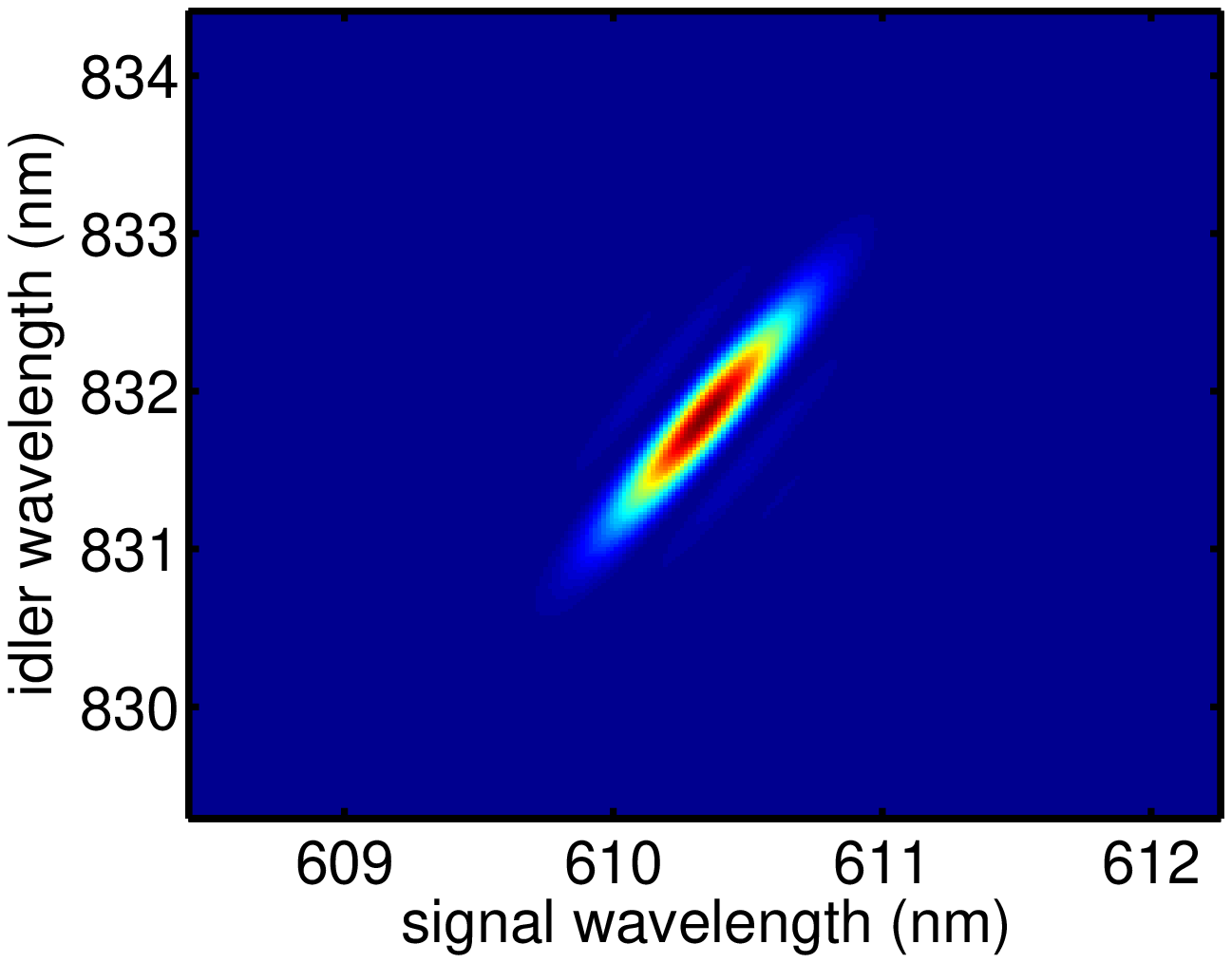}
\end{minipage}
\hspace{0.5cm}
\begin{minipage}[b]{0.4\textwidth}
\centering
(b)
\includegraphics[width=\textwidth]{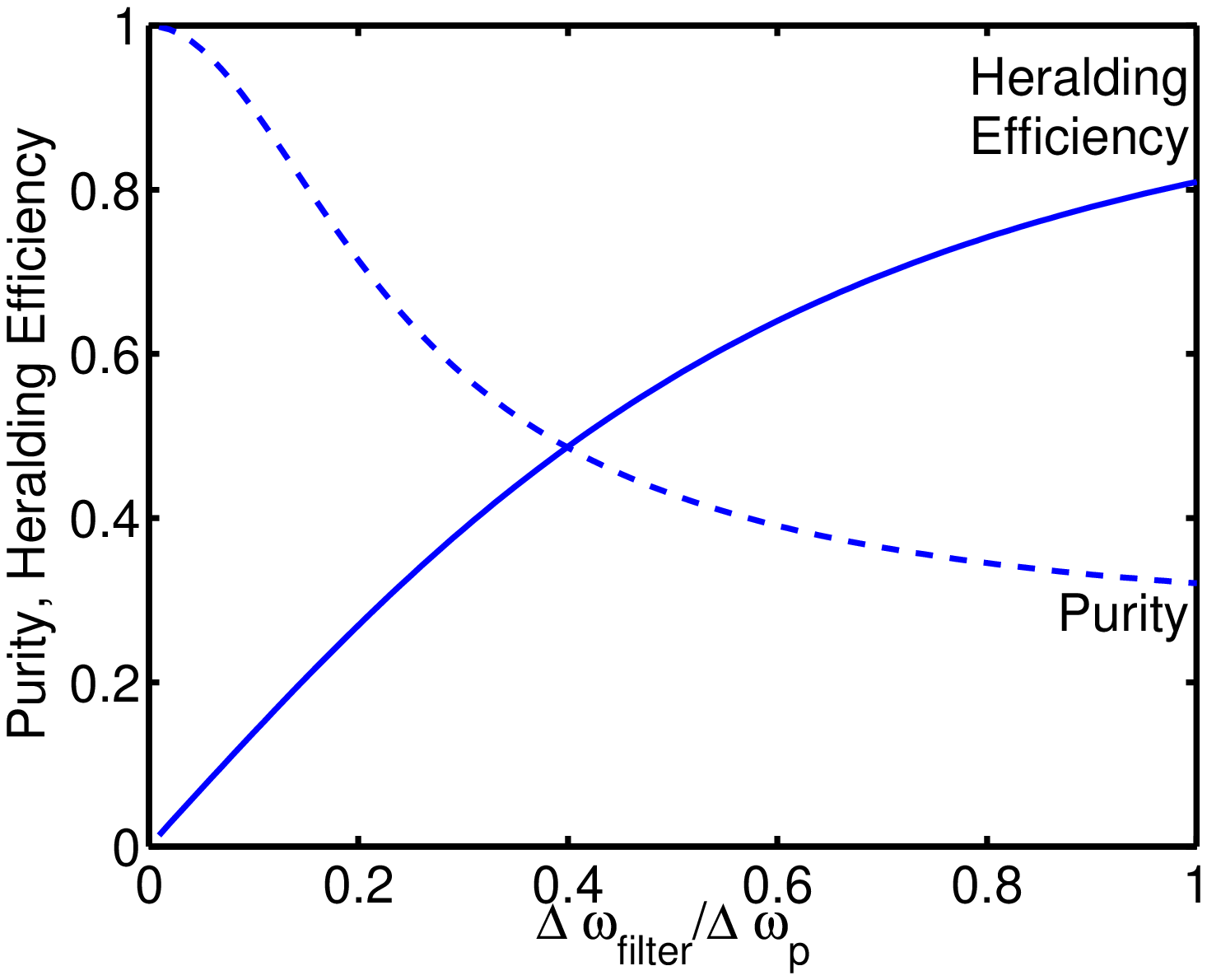}
\end{minipage}
\end{minipage}

\begin{minipage}[b]{1\textwidth}
\begin{minipage}[b]{0.4\textwidth}
\centering
\includegraphics[width=\textwidth]{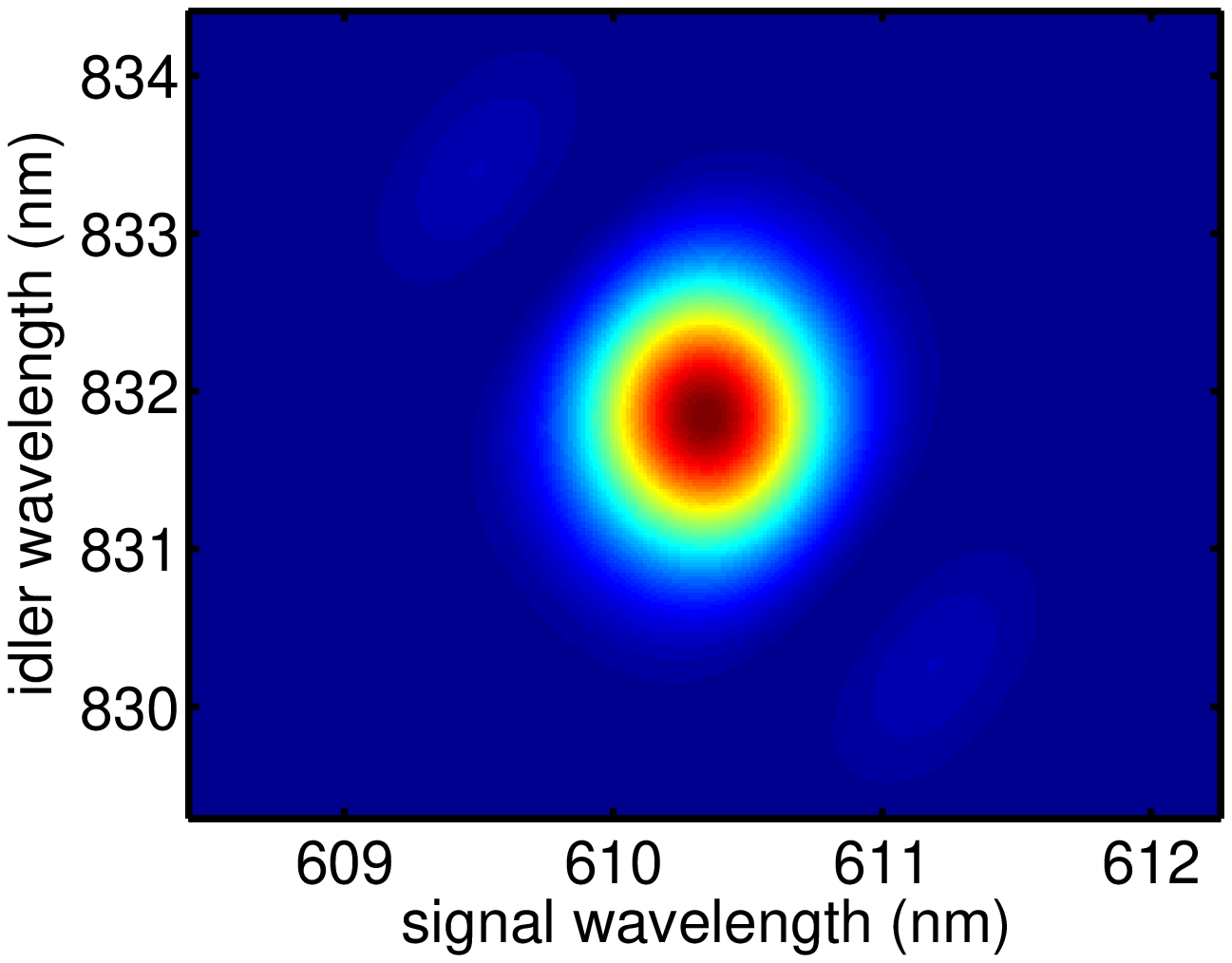}
(c)
\end{minipage}
\hspace{0.5cm}
\begin{minipage}[b]{0.4\textwidth}
\centering
\includegraphics[width=\textwidth]{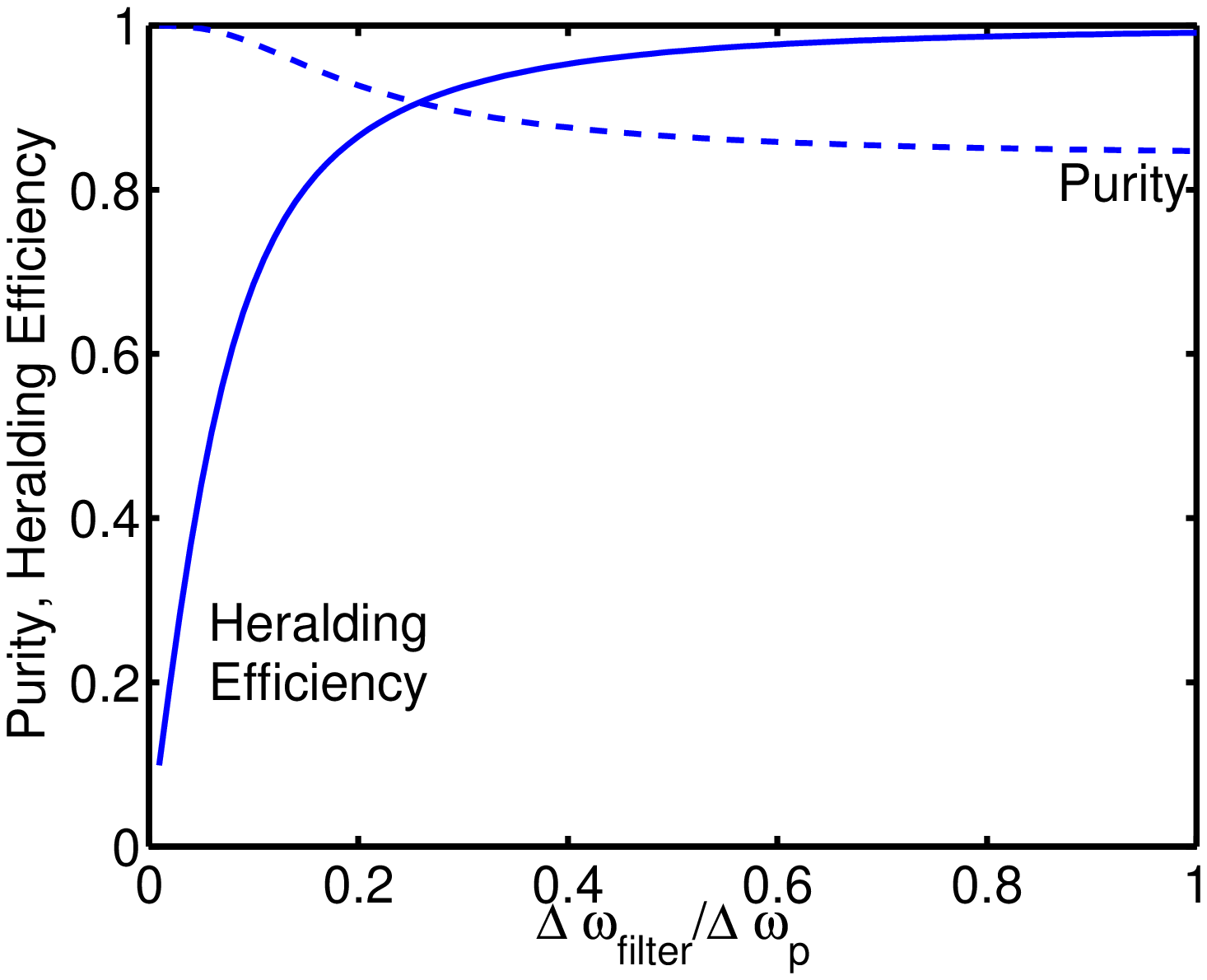}
(d)
\end{minipage}
\end{minipage}
\caption{(a) A typically correlated joint spectrum for a spontaneous four-wave mixing (SFWM) photon-pair source.  (b) Heralding efficiency (solid) and purity (dashed) as a function of Gaussian filter bandwidth (in units of the pump bandwidth) for the correlated case.  (c) Joint spectrum for a SFWM photon-pair source with matched phase matching and pump bandwidths.  (d) Heralding efficiency (solid) and purity (dashed) as a function of Gaussian filter bandwidth (in units of the pump bandwidth) for the matched bandwidths case.}
\label{fig:jointSpec}
\end{figure}

In general, photon pair sources based on SPDC or SFWM produce pairs which display spectral entanglement and this manifests itself in a joint spectral amplitude (JSA) which is not factorable.  A typical example of such a JSA is shown in figure~\ref{fig:jointSpec}(a) where the correlations are clearly visible as correlations in the frequencies of the two photons.  Consequently, although detecting $N$ photons in one mode heralds the presence of $N$ photons in the other, the heralding detectors are typically spectrally non-selective and therefore project these heralded photons into a spectrally mixed state.  The usual technique for removing these spectral correlations is to use narrow spectral filters, but this improvement in purity comes at the expense of decreased heralding efficiency, i.e.~decreased $\eta_{p}$, which therefore also increases the effects of noise in the experiment.

The trade-off between heralding efficiency and spectral purity for the highly correlated joint spectral amplitude of figure~\ref{fig:jointSpec}(a) is shown in figure~\ref{fig:jointSpec}(b).  The purity of the unfiltered heralded signal photon is 27\%. Using spectral filters that only pass a symmetric portion of the joint spectrum dramatically decreases the heralding efficiency of the source so that, for a purity of 95\%, heralding efficiency is limited to 9.7\%.

A promising route to generating heralded pure-state single photons with high efficiency and purity is to directly engineering the nonlinear optical interaction~\cite{GriceWP2001efs, URenAB2003peq, URenAB2005gps, Garay-PalmettK07ppp, MosleyPJ2008hgu, CohenO2009tpg, HalderM2009n2i}. In these approaches, an intense pump is converted into a pair of photons, denoted signal (s) and idler (i) mediated by the nonlinear medium. This process must obey both energy and momentum conservation, which generally leads to spectral and momentum correlations. However, by carefully choosing material parameters and pump configuration, these correlations can be eliminated.  SFWM utilizing birefringent phase matching in optical waveguides offers several advantages over other nonlinear optical sources, such as perfect spatial mode matching with existing photonic circuitry allowing low-loss integration, flexible tuning of the signal and idler central wavelengths through birefringence, and tolerance to waveguide properties~\cite{SmithBJ2009ppg, SollerC2010bvt}.  In this scenario, the pump propagates along the fast axis while the signal and idler photons are produced along the orthogonal slow axis of the waveguide.  The wave-vector mismatch includes a term proportional to the birefringence and pump central frequency. Thus, by adjusting the pump central frequency and birefringence one can tune the central frequencies of the signal and idler photons. To achieve the nearly factorable states required for heralding pure-state single photons with birefringent phase matching, the pump bandwidth must be balanced with the phase matching bandwidth which is determined by the interaction length~\cite{SmithBJ2009ppg, SollerC2010bvt}.  A typical JSA produced under such balanced conditions is shown in figure~\ref{fig:jointSpec}(c).  The correlations are significantly reduced despite the remaining lobes which result from the sinc-like form of the phase-matching function.  Figure~\ref{fig:jointSpec}(d) shows the effect on the heralding efficiency and purity of applying spectral filters to this JSA.  For a purity of 95\%, heralding efficiency in this case is greater than 80\% directly demonstrating the higher intrinsic purity of the source.

\section{Manipulation}

Once pure single photons have been created, they must be manipulated in such a way as to implement, for example, logic gates or a metrology scheme.  Only two components are necessary in order to implement all unitary operations: beam splitters and phase shifters.  Beam splitters can be constructed using evanescent~\cite{MarshallGD2009lww, PolitiA2008swq}, x-couplers~\cite{SmithBJ2009pip}, or multimode interference devices~\cite{PeruzzoA2010mqi} while phase shifters can be implemented by thermo-optic~\cite{SmithBJ2009pip, MatthewsJCF2009mme} or electro-optic~\cite{HerrmannH2010tie} means.  

Evanescent couplers are fabricated by bringing together two waveguides so that the evanescent fields of the two modes overlap, creating a coupling between them.  The field then flips back and forth between the modes as they propagate in parallel.  The rate of this flipping is determined by the overlap of the two modes which is dependent on the separation of the waveguides.  For such a device with a given waveguide separation, the splitting ratio is determined by the length of the coupling region, and the wavelength being used.  They can be produced using either lithographic or direct-write techniques.  X-couplers are produced by physically crossing the waveguides at a shallow angle giving a splitting ratio which is a function of this angle.  Direct write techniques must be used to fabricate x-couplers since in order to produce this shallow angle, sharp features in the guiding structure are required which cannot currently be produced using lithographic techniques.  

A detailed comparison of thermo-optic phase shifters against electro-optic versions is yet to be performed for quantum devices because, although electro-optic phase shifters are common in commercially available classical integrated devices, they have not been demonstrated in an integrated photonic device.  They have several potential advantages over their thermo-optic counterparts, however, most notably a reduced switching time which, if combined with integrated quantum memories, could provide a route to one-way photonic QIP.

For the specific example considered in this paper, it is critical to be able to produce beam splitters with accurate splitting ratios.  The splitting ratios of the two beam splitters shown in figure~\ref{fig:schemSetup}(b) affect both the fidelity of the HB state produced and the measurement scheme which is implemented, limiting the FI that can be obtained.  

\begin{figure}[h!]
\begin{minipage}[b]{0.4\textwidth}
\centering
\includegraphics[width=1\textwidth]{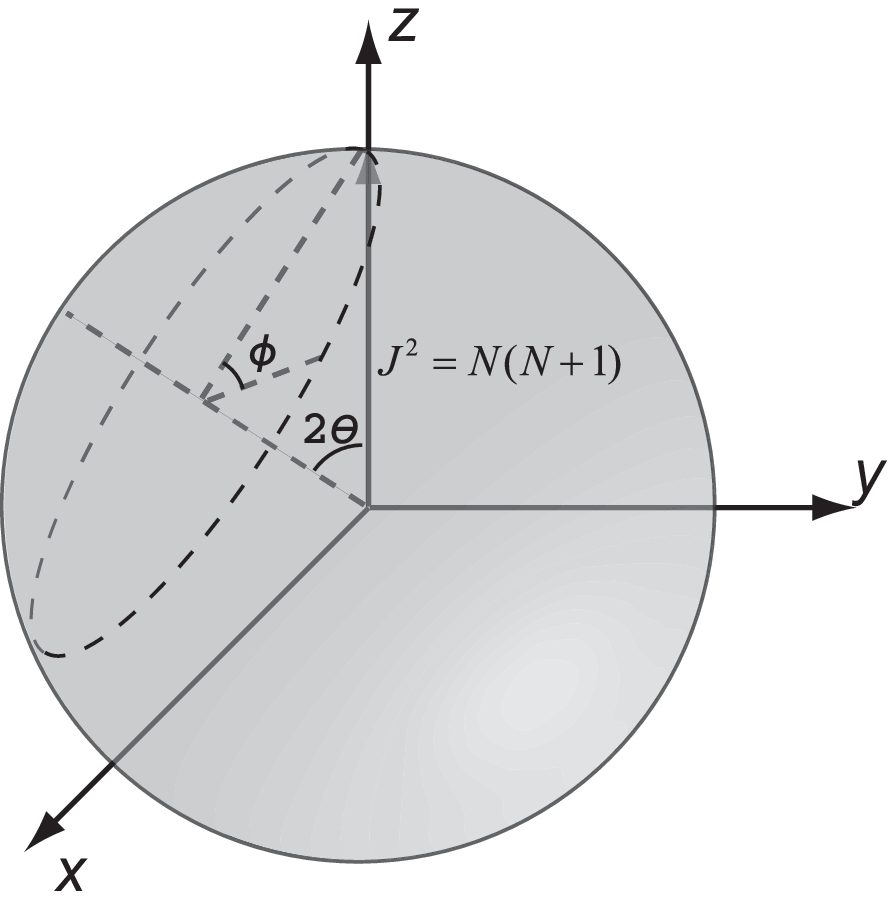}
(a)
\end{minipage}
\hspace{0.5cm}
\begin{minipage}[b]{0.4\textwidth}
\centering
\includegraphics[width=1\textwidth]{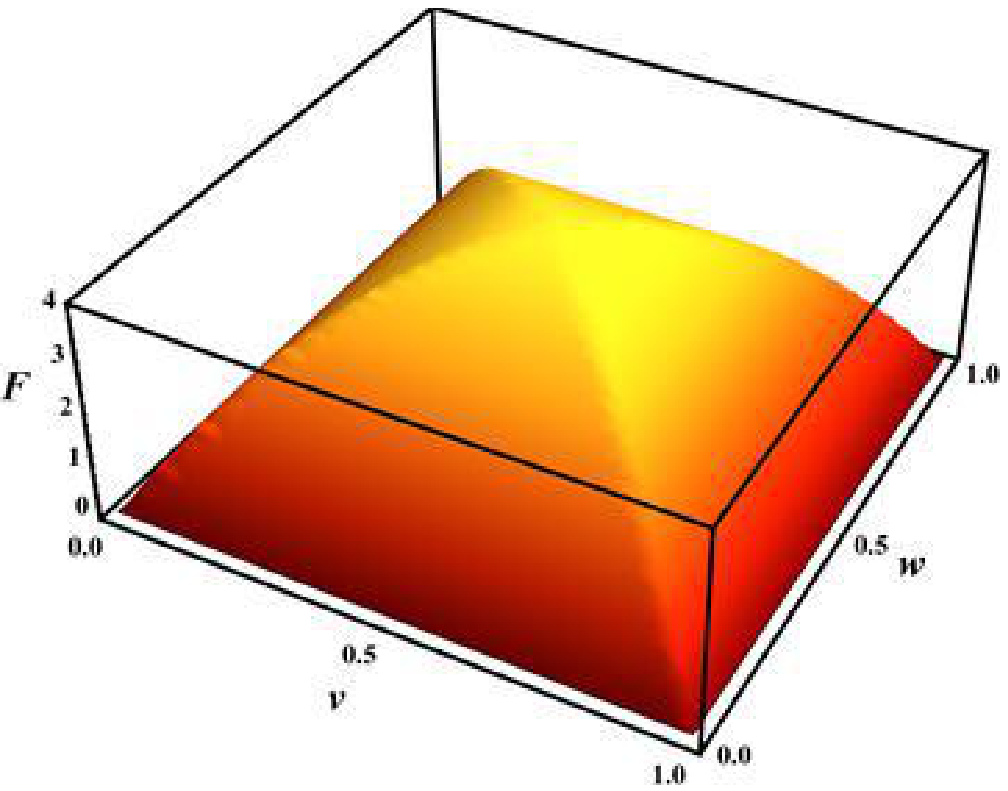}
(b)
\end{minipage}
\caption{(a) Depiction of the effect of varying $v$ and $w$ on the effective phase operator.  (b) The Fisher information of the state generated by $\ket{1}_a\ket{1}_b$ impinging on a beam splitter of reflectivity $v$ when a detection event $\proj{1}\otimes\proj{1}$ occurs after a beam splitter of reflectivity $w$, maximised over the phase $\phi$.}
\label{fig:sphereAndQCRBwithEta}
\end{figure}

In order to show this, we begin by defining the angular momentum operators in the Schwinger representation as:
\ben
J_x = \frac{1}{2}(\adagg b+ a\bdagg),\; J_y = \frac{1}{2i}(\adagg b - a\bdagg),\; J_z = \frac{1}{2}(\adagg a - \bdagg b),\\
J^2 = J_x^2 + J_y^2 + J_z^2.
\een
In a basis defined by the mutual eigenbasis of $J^2$ and $J_z,$ the ideal input state $\ket{N}_a\ket{N}_b$ appears as $\ket{N,0}.$ In this picture, the phase operator is~\cite{SandersBC1997oqm}
\be
P(\phi) = e^{i\phi J_z},
\ee
while a beamsplitter is given by 
\be
B(\theta) = e^{-2i\theta J_x},
\ee
where $\sin^2\theta = v$ is the input beamsplitter transmissivity. The effective phase operator is then given by 
\be
\label{newaxis}
\tilde{P}(\phi) = B(\theta)P(\phi)B(\theta)^{\dag} = e^{i\phi(\cos2\theta J_z - \sin2\theta J_y)},
\ee
where we have used the SU(2) commutation relations of the angular momentum operators. For a 50/50 beamsplitter, $\theta = \pi/4,$ in which case the effective role of the beamsplitter and the phase combined is to effect a rotation about the $-J_y$ axis. In general, this axis of rotation is given by~(\ref{newaxis}), and given that the input state points along the $z$-axis, the radius of rotation shrinks by a factor of $\sin2\theta=2\sqrt{v(1-v)} \leq 1,$ as depicted in figure~\ref{fig:sphereAndQCRBwithEta}(a). The net effect is to suppress the phase picked up by the state $\ket{N}_a\ket{N}_b$ with the beamsplitter ratio $v$.  Since the phase picked up is suppressed by this factor, the precision is also suppressed by the same amount so that the QFI in this case is given by
\be
\label{vQFI}
F_\mathrm{Q} = 4N(N+1)v(1-v).
\ee
This shows that the best attainable precision is provided by a 50/50 input beamsplitter. In that case, we know that this can be attained with another 50/50 beamsplitter at the other end of the interferometer and an $\proj{N}{}\otimes\proj{N}{}$ detection (in the loss-less case). For every other situation, however, there remains the issue of attaining the reduced bound provided by~(\ref{vQFI}).  Allowing for a general beamsplitter at the detection end makes the calculation somewhat involved, so we restrict ourselves to the $\ket{1}_a\ket{1}_b$ case. Then the classical FI, $F$, as a function of the input beamsplitter $v$ and output beamsplitter $w$ is shown in figure~\ref{fig:sphereAndQCRBwithEta}(b), assuming that we have perfect photon-number resolving detectors.

\begin{figure}[h]
\begin{minipage}[b]{0.45\textwidth}
\centering
\includegraphics[width=\textwidth]{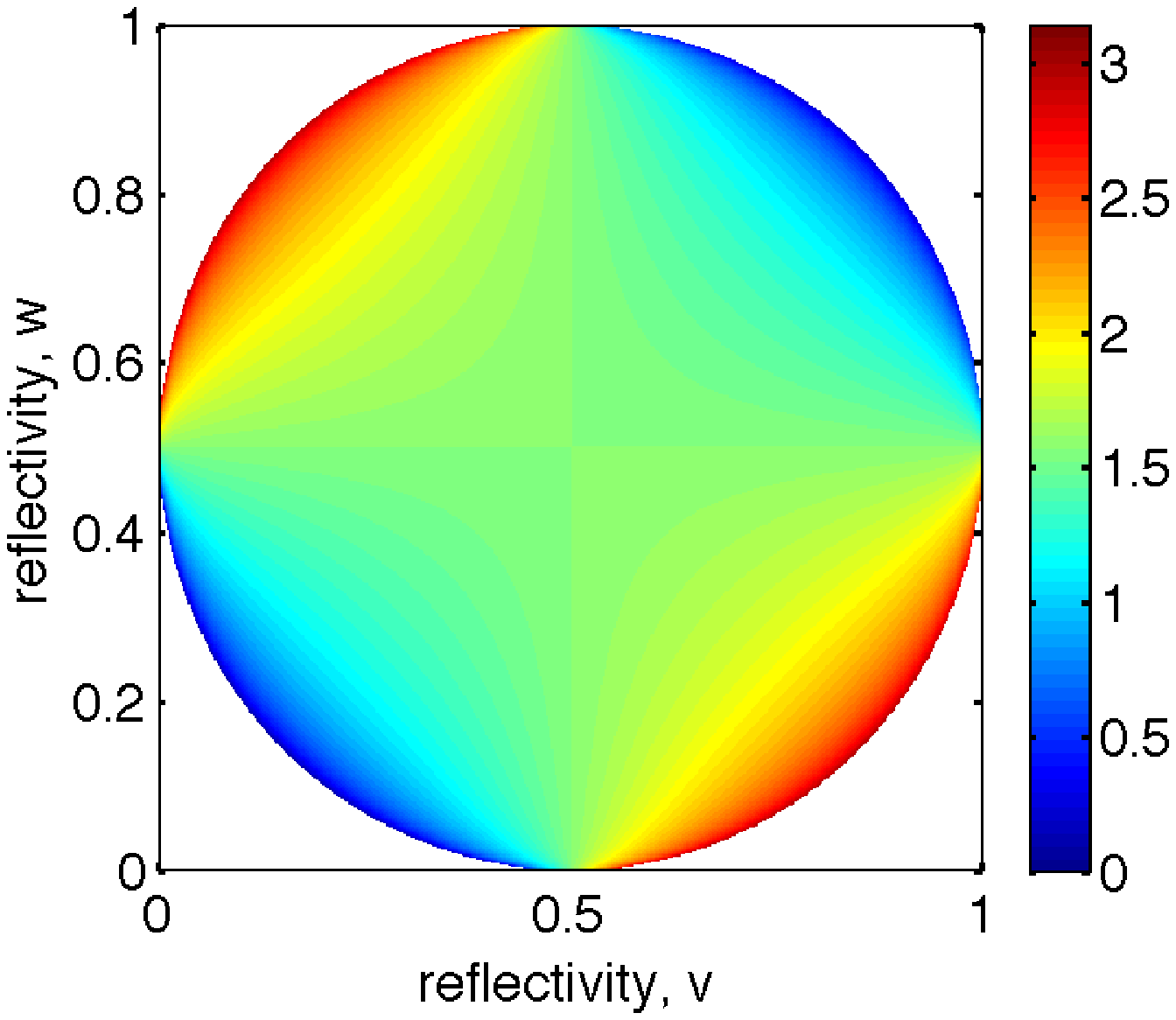}
(a)
\end{minipage}
\hspace{0.5cm}
\begin{minipage}[b]{0.45\textwidth}
\centering
\includegraphics[width=\textwidth]{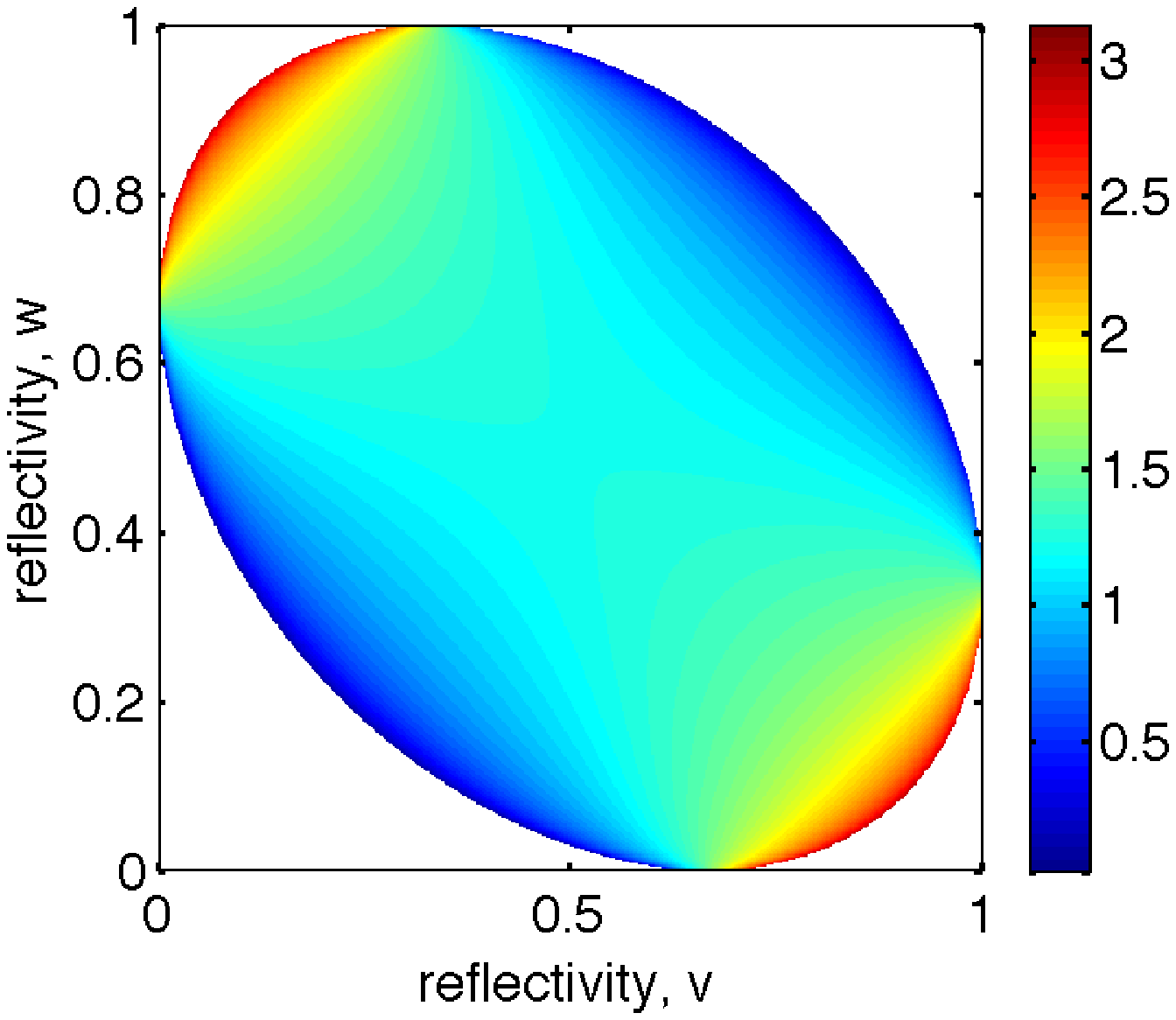}
(b)
\end{minipage}
\caption{(a) The values of $v,w$ for which an effective reflectivity of 1/2 can be reached.   (b) The values of $v,w$ for which an effective reflectivity of 1/3 can be reached.  In both plots the reflectivity combinations in the white areas do not allow tuning to the target effective reflectivity.  The phase required to reach the target effective reflectivity is represented by the colour.}
\label{fig:MZIto50:50}
\end{figure}

In practice, fabrication tolerances result in beamsplitters which do not have the reflectivity they were designed to.  Fortunately, the MZI provides a solution, since the combination of two beam splitters and a phase shifter can be thought of as a single beam splitter whose reflectivity, $v_\mathrm{eff}$ is a function of the applied phase,
\ben
v_\mathrm{eff} = vw + (1-v)(1-w) + 2\sqrt{v(1-v)w(1-w)}\cos{\phi}.
\label{eqn:effectiveRef}
\een  
If both beam splitters were to have a splitting ratio of 1/2 then, by adjusting the phase, the full range of reflectivities from perfectly reflective to perfectly transmissive can be realised.  As the reflectivities of the beam splitters vary away from 1/2, the effective reflectivities which can be reached by tuning the phase is reduced, however this technique dramatically increases the robustness of the system to fabrication tolerances.  Figure~\ref{fig:MZIto50:50} shows the values of the two beam splitters for which reflectivities of 1/2 and 1/3 can be obtained.  In principle, one could even combine four nominally 50:50 beam splitters with three phase shifters to produce a single programmable beam splitter which is tunable over the full range $0 \le v_\mathrm{eff} \le 1$.

\begin{figure}[h]
\input{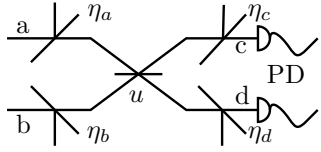}
\caption{Schematic of a device consisting of a single beam splitter of reflectivity $u$ with input and output coupling efficiencies modelled by beam splitters with $\eta_{a},\eta_{b}$ for input coupling transmissivity, and $\eta_{c},\eta_{d}$ for output coupling transmissivity. The spatial modes are labelled a, b, c, d.  Detection is performed by photodiodes (PD).}
\label{fig:bsWithCoupling}
\end{figure}

Since variation in beam splitter reflectivities is inevitable, a way of accurately characterising them is necessary in order to calibrate and predict the behaviour of any device.  Full process tomography becomes extremely complex and unwieldy for large systems so it is necessary to characterise components individually to build up the characteristics of the device.  Here we present a ratiometric characterisation technique which allows measurement of the splitting ratio independently of the input and output coupling efficiencies associated with launching light into and collecting light from the device.  The situation being considered is shown schematically in figure~\ref{fig:bsWithCoupling}.  Although only a single beam splitter is shown, this technique can be applied to any beam splitter embedded in a larger circuit as long as both inputs to and both outputs from the beam splitter can be accessed independently.

Initially, coherent light with intensity $I$ is coupled into mode a.  This light propagates through the circuit and is detected at the outputs, having experienced propagation, coupling, and detection losses. These can be represented by a single transmissivity at the inputs, $\eta_{a}$ and $\eta_{b}$, and a single transmissivity at the outputs, $\eta_{c}$ and $\eta_{d}$.  At the outputs, two intensities are measured, $I_{ac}$ and $I_{ad}$ in modes c and d respectively.  These intensities can be written as a function of the transmissivities, beam-splitter reflectivity, and the initial intensity, 
\begin{eqnarray}
I_{ac} &=& \eta_{a}u \eta_{c}I, \\
I_{ad} &=& \eta_{a}(1-u)\eta_{d}I.
\end{eqnarray}
Similarly, the same light is coupled into mode b, giving
\begin{eqnarray}
I_{bc} &=& \eta_{b}(1-u)\eta_{c}I \\
I_{bd} &=& \eta_{b}u \eta_{d}I.
\end{eqnarray}
Finally, by taking the ratio $r = I_{ad}I_{bc}/I_{ac}I_{bd}$ and solving for $u$, taking the solution which gives $0\leq u\leq1$, we get
\begin{equation}
u = \frac{1}{1+\sqrt{r}}.
\end{equation}

\begin{figure}[h]
\includegraphics[scale=0.5]{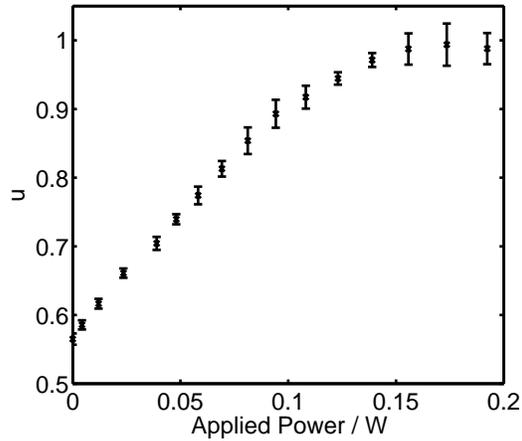}
\caption{The effective reflectivity of a MZI as a function of the power dissipated by a thermo-electric phase shifter.  Characterisation was performed using a ratiometric technique which is independent of coupling losses (see text).}
\label{fig:mziAsBScharacterisation}
\end{figure}

Figure~\ref{fig:mziAsBScharacterisation} shows the results of this technique being applied to a MZI which is acting as a single, programmable beam splitter.  The maximum beam splitter reflectivity reached is 0.993.  The range of applied powers is limited in this case due to the robustness of the particular device used.  Fitting equation~(\ref{eqn:effectiveRef}) to a full fringe would allow the individual beam splitter reflectivities to be obtained, however, this equation is symmetric under exchange of the two reflectivities and so, although the values can be obtained, the ordering is ambiguous.  In order to resolve this ambiguity, we must be able to measure directly one of the beam splitter ratios.  As well as this, the individual inputs and outputs of some beam splitters in QIP circuits are not independently accessible, for example in a CNOT gate where only three beam splitters may be characterised in this way~\cite{LaingA2010hoq}.  

This problem is alleviated by using a camera to collect light scattered out of the mode in the transverse direction.  The outputs of a particular beam splitter can then always be independently accessed, drastically increasing the applicability of the above technique.  In this situation, the light observed at the camera is a proportion of the light propagating in the mode.  As with the ratiometric technique outlined above for output intensities measured with a linear detector, the proportion scattered does not need to be constant for different paths since the ratiometric analysis gives a reflectivity which is independent of the scattering ratio and coupling efficiency.  Consequently, however, the camera used must display a linear response and the image of the output modes must not change position on the camera sensor so that the pixel efficiency is constant when coupling into each input.

\section{Measurement and Detection}

The final stage in any photonic experiment is detection.  High-quality, quantum-limited measurements are an essential requirement for demonstrating explicitly non-classical behaviour in a quantum experiment, and this is characterised by the detectors, the measurement apparatus as a whole, and by the way that apparatus is utilised in the experiment. Quantum-limited detection in photonics experiments is normally carried out using photon counting.  As discussed above in relation to state preparation, in the metrology example, the detection must be very efficient if the quantum device is to out-perform its classical counterpart in terms of overall precision~\cite{Thomas-PeterNL2010rqs, DattaA2010qmi}.  Indeed, even with detector efficiencies close to 100\%, the constraints placed on preparation efficiencies and propagation losses for achieving true quantum-enhanced operation are extremely stringent and have not been demonstrated by any existing experiment.

The standard workhorse of photon-counting experiments is the APD, a binary (``zero/many'') detector which is readily commercially available.  Unfortunately, typical quantum efficiencies for silicon APDs reach at maximum up to 60\% for wavelengths around 800 nm, where they have a peak in sensitivity.  By contrast, efficiencies of up to 98\%~\cite{LitaAE2010sts} can be reached with the recently developed superconducting transition-edge sensors (TES detectors), with the additional advantage that they also provide direct photon-number resolution.  Both APDs and TES detectors are generally operated in a fibre-coupled configuration, making them well-suited to a modular integrated-photonics platform.  A major drawback with the still-emerging TES technology is that the detectors currently have to operate at ultra-cold cryogenic temperatures (typically mK).  While the efficiency is still limited, another way to achieve many of the benefits of photon-number resolution is via time multiplexing~\cite{AchillesD2006dlc, AvenhausM2008pns}.  Time-multiplexed detectors (TMDs) are pseudo-photon-number-resolving detectors which utilise standard, commercial components and can be operated straightforwardly at room temperature.  They work by dividing the input signal into a number of distinguishable temporal modes, which are then monitored by standard APDs.  The scaleability of this technique is largely limited by the dead-time of the APDs, which is typically around 50 ns for current APDs.  This currently leads to large readout-response times and therefore limits the repetition rate at which they are able to operate.  TMDs only provide pseudo-photon-number resolution, because there is still some probability that more than one photon will arrive at an APD in a single temporal mode---they only provide unambiguous photon-number resolution in the limit of many more temporal modes than incoming photons.  These effects can be in many cases mitigated, however, by appropriately characterising the behaviour of the detectors in advance via detector tomography~\cite{LundeenJS2009tqd, FeitoA2009mmt, WorsleyAP2009aee}.

Detector tomography is a general technique which uses weak-coherent input states to probe the detector operation and reconstruct the individual measurement operators representing each possible detector outcome, including their description in the photon-number basis.  To date, this technique has been used to characterise the behaviour of both a standard APD and a TMD~\cite{LundeenJS2009tqd, FeitoA2009mmt}.  An important aspect of detector tomography is that it provides a direct estimate of detection efficiency \cite{WorsleyAP2009aee} which, as we have already discussed, is key to understanding and interpreting observations of quantum effects in experiments (e.g.~metrology), particularly in the presence of loss.

A common task that is often necessary in the building and characterisation of photonic circuits for quantum applications is quantum state tomography~\cite{SmitheyDT1993mwd, JamesDFV2001mq, HradilZ2006bts}.  In order to understand the performance of the quantum device, it is important to be able to estimate the form of the input quantum state upon which that performance depends.  This is equally true for integrated photonics applications.  Traditionally, photonic experiments rely on post-selection, where the state is assumed to be within a specific photon-number subspace.  This means that loss in the measurement device used to perform state tomography can, for the most part, be ignored.  Unfortunately, however, this is not sufficient in many situations, such as the metrology example that we have considered here.  In this case, phase information can generally be acquired in some degree from all photon-number subspaces, particularly in the case of ``classical'' coherent input states, against which any quantum device must be measured to determine whether it improves upon the classical limit.  The ability to acquire some level of phase information from lower photon-number terms is a signature of loss-tolerant probe states, such as the optimally loss-tolerant state~\cite{DornerU2009oqp, KacprowiczM2010eqe} and the HB states~\cite{HollandMJ1993ido}.  Unusually, this is not the case with the N00N state, which is an indication of its fragility to loss.  However, because of the effects of loss, even in this case it is critical to characterise the complete input state, including the contributions from all photon-number subspaces, in order to obtain a genuine estimate of the performance of the quantum device.  Once the complete density matrix of the input state is known, this can be used to calculate the QCRB, which defines the usefulness of the prepared state for phase estimation.  In order to reconstruct the complete density matrix of a quantum state, two key experimental ingredients are required: firstly, a heralded source of input states to probe the contribution from the vacuum subspace, and secondly, a measurement apparatus which provides some level of photon-number resolution, so that different Fock subspaces can be interrogated.

\begin{figure}[h]
\input{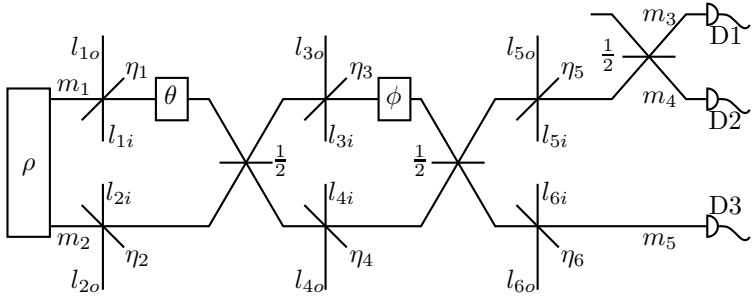}
\caption{A typical setup used to perform state tomography on $\rho$.  $\rho$ is assumed to be a single polarisation photonic state in two spatial modes.  A MZI and two phase shifters allow a tomograpically complete set of measurements to be made.  Detection is performed by avalanche photodiodes (D1-D3).  The spatial modes which lead to detectors are labelled $m_{1}$-$m_{5}$ while loss modes are labelled $l_{1i}$-$l_{6i}$ and $l_{1o}$-$l_{6o}$.}
\label{fig:tomographySetup}
\end{figure}

In general, a state tomography experiment can be described by a set of $N_{\alpha}$ detection apparatus settings, labelled here with $\alpha$, each of which has associated with it a set of $N_{\gamma}$ outcomes, labelled here with $\gamma$ so that $N_{\alpha}N_{\gamma}$ measurements are implemented.  In photonic experiments, the outcomes are usually a particular combination of physical detectors firing simultaneously (``in coincidence'').  The measurements implemented are modelled by a set of positive operator value measure (POVM) elements $\{\hat{\Pi}_{\gamma}\}_{\alpha}$.  In this scheme, the outcomes for each setting are complete so that for every $\alpha$, $\sum_{\gamma}\hat{\Pi}_{\gamma}=\mathbbm{1}$.  The settings must be chosen so that the POVM elements span the space of states, allowing reconstruction of any state in the space.  In general there are optimal ways to choose both the settings and the amount of time spent measuring at each setting, which is sometimes referred to as optimal experiment design~\cite{deBurghMD2008cms, AdamsonRBA2010iqs, LingA2006eps, NunnJ2010oed}.

Once the settings have been chosen, the measurements are performed, usually with a fixed time spent at each $\alpha$.  In this time, the number of events corresponding to each outcome $\gamma$ are recorded, giving a set of counts $\{n_{\gamma}\}_{\alpha}$.  The probability that any state, represented in density matrix form as $\rho$, could produce the observed counts is then given by the likelihood function, 
\begin{equation}
\mathcal{L} = p(\{n_{\alpha\gamma}\}|\rho),
\end{equation}
where $\{n_{\alpha\gamma}\}$ denotes the set of all measured counts at all settings.  We discuss here the most common state reconstruction technique, known as maximum likelihood estimation, where the estimated $\rho$ is determined by maximising the value of the likelihood function~\cite{JamesDFV2001mq, HradilZ2006bts}.

The exact form of $\mathcal{L}$ is dependent upon the system generating the state.  With heralded state generation, the number of times that the state has been prepared for each setting, $n_{\alpha}$, is known so that for each herald signal a result is collected.  The number of heralds limits the values of $n_{\gamma}$ so that $\mathcal{L}$ is most appropriately described by a multinomial distribution,
\begin{equation}
\mathcal{L} = \prod_{\alpha\gamma}C_{\alpha}(\{n_{\gamma}\}_{\alpha})p_{\alpha\gamma}^{n_{\alpha\gamma}},
\end{equation}
where $C_{\alpha}$ is a multinomial factor which accounts for the number of ways the observed $n_{\gamma}$ could have occured for each setting and $p_{\alpha\gamma} = tr(\rho\hat\Pi_{\alpha\gamma})$.  Since $n_{\alpha}$ is known for each setting, only $N_{\gamma}-1$ measured counts, or their corresponding probabilities, are independent variables in the maximum likelihood optimisation.  This can be expressed as the constraint equation $\sum_{\gamma} n_{\alpha\gamma} = n_{\alpha}$.  This maximisation can be performed directly using an iterative technique~\cite{HradilZ2006bts}.

If $n_{\alpha}$ is large and $p_{\alpha\gamma}$ is small, the multinomial form can be approximated by a product of $N_{\alpha}(N_{\gamma}-1)$ Gaussian distributed variables,
\begin{equation}
\mathcal{L} = \prod_{\alpha\gamma}\frac{1}{\sqrt{2\pi\sigma_{\alpha\gamma}^{2}}}\exp\{-\frac{(n_{\alpha\gamma}-n_{\alpha}p_{\alpha\gamma})^2}{2\sigma_{\alpha\gamma}^{2}}\},
\end{equation}
where $\sigma_{\alpha\gamma}=n_{\alpha}p_{\alpha\gamma}(1-p_{\alpha\gamma})\approx n_{\alpha}p_{\alpha\gamma}$ for small $p_{\alpha\gamma}$.  Typical current devices exhibit high loss, so that the vacuum component of any heralded state will contain a large population.  While this might be expected to take us out of the regime of small $p_{\alpha\gamma}$, we are still free to choose which outcome probability is the ``dependent variable'' for each measurement setting.  Using the corresponding constraint equation, we therefore choose to eliminate the vacuum term, leaving only the lower probability terms which still satisfy the small $p$ approximation.  Expressing the likelihood function in this form ultimately allows one to express the problem in a weighted least-squares form which can be converted into a semi-definite program so that the well developed tools of convex optimisation can be employed~\cite{BoydS2004co}.

A necessary part of this optimisation is knowledge of the POVM elements which model the particular outcomes observed.  Since we wish to be able to reconstruct all photon number subspaces, we need measurements which access all photon number subspaces.  In the case of a perfectly efficient, perfectly photon number resolving device, it can be precisely known which photon number subspace is being projected onto.  In the presence of any loss, however, even photon number resolving detectors project onto multiple photon number subspaces since registering a single photon could be due to, for example, two photons where one is lost.  The exact value of the loss in the measurement device specifies the mixture of photon number subspaces being projected onto and hence must be known if a state is to be accurately reconstructed.  As an example, we will consider the device shown in figure~\ref{fig:tomographySetup} which provides tomographic measurements suitable for characterising the HB(1) state.  The detectors used are perfect APDs which have a POVM set of $\{\proj{0}, \mathbbm{1}-\proj{0}\}$ meaning that they either don't fire and therefore project onto the vacuum, or they fire in which case at least one photon was present but it is not known how many.  This device is insensitive to coherences between components in different photon number subspaces.  Fortunately, however, these coherences cannot affect the measurement statistics in our system, or indeed in any system comprised of imperfect state preparation, lossy linear optical interactions, and inefficient detection.  This is true because all linear optical components conserve photon number by definition (i.e.\ they operate only within each photon number subspace), loss is always incoherent, and photon counting detectors project onto incoherent mixtures of components from within these different subspaces.  A state reconstructed in this way therefore provides sufficient information to predict all measurement outcomes of any subsequent system acting on that state using linear optical interactions and photon counting measurement. 

For the system in figure~\ref{fig:tomographySetup}, there are five possible outcomes: no clicks; a click at either of detector 1 (D1) or detector 2 (D2); a click at detector 3 (D3); a click at D1 and D2; a click at D1 and D3 or D2 and D3.  These are labelled $\hat\Pi_{\gamma}^\mathrm{(out)}$ where $\gamma=1$ to 5.  Losses in the system are modelled by beam splitters with reflectivities $\eta_{1}$ to $\eta_{6}$.  Detector inefficiencies are incorporated into the losses directly before the detectors ($\eta_5$ and $\eta_6$)~\cite{AchillesD2004pdu}.  The modes leading to detectors are labelled with $m_{1}$ to $ m_{5}$ while the loss modes are labelled $l_{1o}$ to $l_{6o}$ for the ``outer'' modes and $l_{1i}$ to $l_{6i}$ for the ``inner'' modes.  Typically, the projectors for the detectors are back propagated and expressed in terms of projectors at the input of the measurement device and the $\eta_{i}$ are set to 1.  If losses are considered, however, one must back propagate both the click projectors for the APDs and identity projectors for each of the outer loss modes since they are not monitored.  The output POVM set is given by the projectors
\begin{eqnarray}
\fl \hat\Pi_{1}^\mathrm{(out)} = \mathbbm{1}_{l_{1o}}\otimes...\otimes\mathbbm{1}_{l_{6o}}\otimes\proj{0}_{m3}\otimes\proj{0}_{m4}\otimes\proj{0}_{m5}, \\
\fl \hat\Pi_{2}^\mathrm{(out)} = \mathbbm{1}_{l_{1o}}\otimes...\otimes\mathbbm{1}_{l_{6o}}\otimes\left[\left(\mathbbm{1}-\proj{0}_{m3}\right)\otimes\proj{0}_{m4} +\proj{0}_{m3}\otimes\left(\mathbbm{1}-\proj{0}_{m4}\right)\right] \nonumber\\
\otimes\proj{0}_{m5}, \\
\fl \hat\Pi_{3}^\mathrm{(out)} = \mathbbm{1}_{l_{1o}}\otimes...\otimes\mathbbm{1}_{l_{6o}}\otimes\proj{0}_{m3}\otimes\proj{0}_{m4}\otimes\left(\mathbbm{1}-\proj{0}_{m5}\right), \\
\fl \hat\Pi_{4}^\mathrm{(out)} = \mathbbm{1}_{l_{1o}}\otimes...\otimes\mathbbm{1}_{l_{6o}}\otimes\left(\mathbbm{1}-\proj{0}_{m3}\right)\otimes\left(\mathbbm{1}-\proj{0}_{m4}\right)\otimes\proj{0}_{m5}, \\
\fl \hat\Pi_{5}^\mathrm{(out)} = \mathbbm{1}_{l_{1o}}\otimes...\otimes\mathbbm{1}_{l_{6o}}\otimes\left[\left(\mathbbm{1}-\proj{0}_{m3}\right)\otimes\proj{0}_{m4} +\proj{0}_{m3}\otimes\left(\mathbbm{1}-\proj{0}_{m4}\right)\right] \nonumber \\
\otimes\left(\mathbbm{1}-\proj{0}_{m5}\right).
\end{eqnarray}

By expanding the photon-number modes as $\proj{n} = 1/n! a^{\dag n} \proj{0} a^n$, these ideal POVM elements can then be propagated backwards through the lossy circuit to determine the complete POVM elements for the overall measurement apparatus, which we represent by $\hat\Pi_{\alpha\gamma}^\mathrm{(in)}$.  These back-propagated projectors, however, can be dramatically simplified using \emph{a priori} knowledge of the system, that is, that the input loss modes $l_{1i}$ to $l_{6i}$ contain only the vacuum, and that the state $\rho$ contains, for example, no more than two photons.  The back-propagated projectors can therefore be \emph{conditioned} using this information, which can be summarised by the following projector:
\begin{equation}
\fl \hat P = (\proj{0}+\proj{1}+\proj{2})_{m1}\otimes(\proj{0}+\proj{1}+\proj{2})_{m2}\otimes\proj{0}_{l1i}\otimes...\otimes\proj{0}_{l6i}.
\end{equation}
This allows us to calculate the final measurement POVM elements according to:
\begin{equation}
\hat\Pi_{\alpha\gamma} = \hat P \hat\Pi_{\alpha\gamma}^\mathrm{(in)} \hat P.
\end{equation}
We note that when the transmissivities $\eta_{1}$ to $\eta_{4}$ are pair-wise symmetric so that $\eta_{1}=\eta_{2}$ and $\eta_{3}=\eta_{4}$, the situation is further simplified, because the total population in the loss modes is not a function of the phases $\theta, \phi$ since every possible path to the detectors experiences the same loss regardless of the phase settings.  This approximation corresponds to the standard bulk-optics scenario where loss does not typically depend on the occupied mode, for example, in a free-space polarization-based interferometer where loss is not normally polarization dependent.

\section{Conclusion}
Integrated photonics is one of several extremely promising technologies whose aims are to allow the exploration of the fundamental properties of quantum mechanics and the application of these phenomena to real tasks.  Although much progress has been made in investigating the individual modules necessary to perform integrated experiments, there has been little work on managing the interfaces between them in order to achieve full integration.  In particular, the effects of loss in quantum devices, which differ significantly from the effects of loss in classical devices, have been largely overlooked to date although they are now the single largest hurdle confronting the field.  In this paper we have explored a modular approach to integration and enumerated the requirements on each of state generation, manipulation, and detection in the context of quantum-enhanced metrology.  We have argued that the control of modes is key, both in order to optimise each individual module, and to optimise the interfaces.   

We have discussed sources based on spontaneous four-wave mixing and the necessary considerations which enable pure, single photons to be generated with high heralding efficiencies in a transverse spatial mode that is well matched to existing photonic circuit technology.  We have shown how managing the spectral mode of the generated photons is critical to maximising the heralding efficiency, demonstrating the trade off between this and state purity for a typically correlated source.  We have also discussed how to construct programmable devices which are robust to fabrication imperfections and demonstrated a method that allows the characterisation of individual beam splitters, even when they are embedded within a complex circuit.  Finally, we have enumerated some of the possibilities for integrated detectors and introduced a framework which describes how to perform state tomography of a heralded state, allowing reconstruction of all photon number sub-spaces rather than post-selecting on a particular photon number sub-space.  This framework ultimately allows the utility of a state for quantum-enhanced metrology to be assessed through the quantum Cram\'er-Rao bound.

\ack
We would like to acknowledge K. Banaszek, C. Silberhorn, U. Dorner and R. Demkowicz-Dobrzanski for fruitful discussions which have informed the progress of this work.  This work has been supported by the European Commission under the Integrated Project Quantum Interfaces, Sensors, and Communication based on Entanglement (Q-ESSENCE), the US European Office of Aerospace Research (FA8655-09-1-3020), the Engineering and Physical Sciences Research Council (EPSRC) (grant EP/H03031X/1), the Royal Society, and also by FONCICYT project  94142.

\section*{References}
\bibliographystyle{unsrt}
\bibliography{abbreviations,integratedPhotonicSensing}

\end{document}